# A continuous metal-insulator transition driven by spin correlations

Yejun Feng[1,2,*], Yishu Wang[2,3], D. M. Silevitch[2], S. E. Cooper[1], D. Mandrus[4,5], Patrick A. Lee[2,6], T. F. Rosenbaum[2,*]

[1]Okinawa Institute of Science and Technology Graduate University, Onna, Okinawa 904-0495, Japan

[2]Division of Physics, Mathematics, and Astronomy, California Institute of Technology, Pasadena, California 91125, USA

[3]The Institute for Quantum Matter and Department of Physics and Astronomy, The Johns Hopkins University, Baltimore, Maryland 21218, USA

[4]Department of Materials Science and Engineering, University of Tennessee, Knoxville, Tennessee 37996, USA

[5]Materials Science and Technology Division, Oak Ridge National Laboratory, Oak Ridge, Tennessee 37831, USA

[6]Department of Physics, Massachusetts Institute of Technology, Cambridge, MA, USA

*Corresponding authors. Email: yejun@oist.jp, tfr@caltech.edu.

**Metal-insulator transitions involve a mix of charge, spin, and structural degrees of freedom, and when strongly-correlated, can underlay the emergence of exotic quantum states. Mott insulators induced by the opening of a Coulomb gap are an important and well-recognized class of transitions, but insulators purely driven by spin correlations are much less common, as the reduced energy scale often invites competition from other degrees of freedom. Here we demonstrate a clean example of a spin-correlation-driven metal-insulator transition in the all-in-all-out pyrochlore antiferromagnet $Cd_2Os_2O_7$, where the lattice symmetry is fully preserved by the antiferromagnetism. After the antisymmetric linear magnetoresistance from conductive, ferromagnetic domain walls is carefully removed experimentally, the Hall coefficient of the bulk reveals four Fermi surfaces, two of electron type and two of hole type, sequentially departing the Fermi level with decreasing temperature below the Néel temperature, $T_N$. Contrary to the common belief of concurrent magnetic and metal-insulator transitions in $Cd_2Os_2O_7$, the charge gap of a continuous metal-insulator transition opens only at $T$~10K, well below $T_N$=227K. The insulating**

**mechanism resolved by the Hall coefficient parallels the Slater picture, but without a folded Brillouin zone, and contrasts sharply with the behavior of Mott insulators and spin density waves, where the electronic gap opens above and at $T_N$, respectively.**

Unlike insulators or semiconductors derived from simple metals such as sodium and lithium [1,2], metal-insulator transitions in correlated-electron systems reside outside the paradigm of single-electron band structure. Factors such as reduced dimensionality and randomness [3, 4] enrich the description of the critical behavior, with possible separation of spin and charge, and deep connections to exotic states such as high $T_c$ superconductivity and quantum spin liquids [5]. In the limit of large electronic correlations, the starting point for discussion is usually the opening of the Hubbard gap [6], with antiferromagnetic order a subsidiary effect.

There exist several major experimental challenges in establishing a convincing example of a metal-insulator transition driven by spin correlations [7]. First, the spin-correlation energy is typically much smaller than the direct Coulomb interaction, often at a scale comparable to that of the structural modifications induced by the antiferromagnetic order through magnetostrictive effects and symmetry changes, leading to a chicken-and-egg conundrum between ascribing the insulating transition to the lattice or the magnetism [8, 9]. It is thus preferable to search for candidate systems in which the magnetism would preserve the crystalline symmetry. Materials that demonstrate all-in-all-out (AIAO) antiferromagnetic order on a pyrochlore lattice, with all four spins on the corner of a tetrahedron pointing either towards or away from the center, meet this criterion. The AIAO order induces no symmetry-breaking magnetostriction, and causes an isotropic expansion of the cubic unit cell by a minimal $\Delta a/a \sim 10^{-5}$, only becoming experimentally resolvable when $T_N$ drops below 40 K [10, 11]. For AIAO order with $T_N > 100$ K, this overall magnetostriction is fully camouflaged by the thermal lattice contraction [10], which broadens bandwidths but introduces no band splitting. The family of AIAO antiferromagnets thus provides a highly desirable model system to investigate spin correlations with little lattice interference.

Identifying an AIAO system with a metal-insulator transition represents the next experimental challenge. Accompanying the AIAO order, many 5$d$ oxides, such as $R_2Ir_2O_7$ ($R$= Eu, Sm, and Nd) and $Cd_2Os_2O_7$ (Ref. [12] and references in [11, 13, 14]), also demonstrate a change of temperature dependence in the resistivity at $T \sim T_N$ [9, 13, 15, 16]. However, the resistive behavior of $R_2Ir_2O_7$ ($R$=Nd, Sm, Eu), especially in the paramagnetic phase, is often inconsistent and raises concerns about their intrinsic, disorder-free behavior [13, 17, 18]. $Cd_2Os_2O_7$ has presented consistent behavior in both the electron correlation and metal-insulator transition. Samples from several groups [9, 15, 16] always demonstrate magnetic ordering at $T_N$=225-227K, metallic behavior above $T_N$, and a three-to-four-decade rise of the resistivity for $T < T_N$ in the best samples. This repeatability from crystal to crystal likely indicates a low level of disorder because of the 2+/5+ valence condition of Cd and Os ions as well as the chemical transport growth procedure at low temperature [9, 15, 16].

There exists an additional challenge arising from complications in modeling and understanding the transport data in many 5$d$ AIAO antiferromagnets due to the intrinsically conductive and highly coercive ferromagnetic domain walls [9, 19-21]. Ferromagnetic domain walls introduce a Zeeman shift in the metallic paramagnetic band structure, but they are not expected to gap the Fermi surface like the antiferromagnetic bulk, and they are expected to remain metallic down to $T = 0$. As was pointed out recently [21], the highly coercive metallic ferromagnetic domain walls generate antisymmetric linear magnetoresistance (MR) of the same functional form as the Hall resistance. Moreover, the antisymmetric linear MR is detectable in Hall channels due to distorted current paths through the domain walls [21]. This effect is likely the root cause of the widely varying Hall coefficient reported in the literature for $Cd_2Os_2O_7$ [9, 16], as the standard procedure for extracting Hall resistance through antisymmetrization with respect to magnetic field direction leads to erroneous results.

Here we present high-fidelity resistivity and Hall coefficient measurements on single crystal $Cd_2Os_2O_7$, after employing an intricate procedure to eliminate the influence of conductive ferromagnetic domain walls. Unlike the common understanding of a concurrent metal-insulator transition with the AIAO magnetic transition at $T_N$, our results reveal that $Cd_2Os_2O_7$ is metallic for a broad temperature range below $T_N$, despite an increasing resistivity with decreasing $T$. It only becomes an insulator at $T_{MIT}\sim 10$ K, when four sets of Fermi surfaces have sequentially left the Fermi level to open a true electronic gap. This large separation in temperature for spin order ($T_N$) and the charge gap ($T_{MIT}$), with $T_N \gg T_{MIT}$, unambiguously establishes spin ordering as the driving force in $Cd_2Os_2O_7$'s metal-insulator transition. Our methodology in separating the Hall behavior of the bulk from the influences of the domain walls should provide a generic approach to parsing spin and charge effects in correlated antiferromagnetic insulators with metallic domain walls.

Our single-crystal $Cd_2Os_2O_7$ samples demonstrate a monotonic rise of resistivity over three decades (~3000×) from $T_N$ to 1.8 K at zero-field (Fig. 1), consistent with the best samples reported in the literature [9, 15, 16]. Instead of using bar-shaped samples for both resistivity and Hall measurements, our key approach is to utilize van der Pauw (vdP) configuration of electrical lead placement on plate-shaped samples ([22], Schematics in Figs. 1, 2a). This choice of putting four leads on equal footing aims to comprehensively evaluate the effects of electrical current paths which constantly redistribute between conductive domain walls and an increasingly insulating bulk as the temperature decreases. As demonstrated by our samples, the vdP ratio, defined as $R_{vdP2}/R_{vdP1}$ at zero field, stays constant only in the paramagnetic phase above $T_N$. Below $T_N$, the vdP ratios of both samples have a strong yet continuous temperature dependence (Fig. 1 inset). The resistivity in Fig. 1 is calculated from $R_{vdP2}$ and $R_{vdP1}$ at each $T$ according to the standard vdP procedure [22], an issue we will revisit below.

The highly coercive, field-independent magnetization **M** of the conductive ferromagnetic domain walls necessarily introduces antisymmetric linear MR [21] that makes the Hall resistances of the two reciprocal channels, $R_{12,34}(\mathbf{H}, \mathbf{M})$ and $R_{43,12}(\mathbf{H}, \mathbf{M})$, have different linear slopes with $H$ (Fig. 2a). Due to the cubic symmetry of AIAO order, it is not possible to create a single antiferromagnetic domain and remove domain walls altogether by field cooling. The separation of galvanomagnetic responses of the bulk domains and the domain walls is instead carried out by introducing a variable **M**($\phi$) through field-cooling along 24 angular directions $\phi$ within the sample surface plane (Fig. 2, Methods, and Ref. [21]). Here we first examine $\phi$-dependences of two reciprocal Hall resistivity channels, the vdP ratio, and resistivity $\rho(H=0)$, at temperatures 195K, 30K, and 1.8K (Fig. 2), where the ferromagnetic domain walls have different levels of influence as gauged by the conductance and the bulk Hall coefficient.

At all three temperatures (Figs. 2b-2d), the average slopes of two Hall resistivity channels are always $\phi$-independent, while the resultant $\phi$-dependent components behave differently. At 195K and 30K, the $\phi$-dependent Hall resistivity slopes of the reciprocal channels follow two constraints, as they are (1) of opposite sign at each $\phi$, canceling out for the $\phi$-independent average, and also (2) identical at $\phi$ and $\phi+\pi$ respectively, leading to identical $\phi$-averaged resistivity slopes of both Hall channels. From previous analysis [21], constraint (1) reflects voltage-current reciprocity, while constraint (2) manifests Onsager's reciprocity relation with regard to ferromagnetic domain wall **M**. At 1.8K, the $\phi$-dependent Hall resistivity slopes (Fig. 2d) behave differently. Only constraint (1) is sustained, while constraint (2) can not be satisfied by any choice of the $\phi$-independent components. The result is that the $\phi$-averaged resistivity slopes of individual Hall channels are no longer necessarily equal to each other and to the $\phi$-independent component.

The breakdown of constraint (2) seemingly implies a violation of Onsager's reciprocity relation between $\phi$ and $\phi+\pi$ states. However, the established Onsager's reciprocity in our system hinges on the inverse relationship between domain wall magnetizations **M**($\phi$) and **M**($\phi+\pi$) through *separate* field-cooling processes at $\phi$ and $\phi+\pi$. Their inverse relationship is generally robust in that both $\rho$ and the vdP ratio at zero field demonstrate a $\pi$-periodicity at all temperatures (Figs. 2b-2d). At 1.8K, the conduction patterns at $\phi$ and $\phi+\pi$ are similar enough to be reproducibly differentiated from those at neighboring $\phi$ positions, judging by $\rho$ and the vdP ratio (Figs. 2d). Nevertheless, because of the separate cooldowns, there exist differences beyond an inversion between **M**($\phi$) and **M**($\phi+\pi$). With the increasingly insulating bulk, the electrical current is more concentrated along a fraction of the domain walls, as reflected by the dramatically oscillating vdP ratio from 3:1 to 1:4 (Fig. 2d). While the entire sample volume brings better averaged galvanomagnetic behavior and demonstrates Onsager reciprocity (Fig. 2c), probing only a small number of domain walls enhances the relative difference between $\mathbf{M}(\phi)$ and $\mathbf{M}(\phi+\pi)$. Even at a fixed $\phi$, multiple field-coolings can lead to significant difference in individual Hall channels at 1.8K (Fig. 2d), despite the consistency at 30K (Fig. 2c). We note that the assumed uniform medium for the vdP technique [22] is justified *a posteriori* at 195K and 30K, as the calculated $\rho(\phi)$ varies within ±0.1% and ±3%, respectively. Constraint (2) remains satisfied (Figs. 2b, 2c). At 1.8K, $\rho(\phi)$ calculated from the vdP formula [vdP1958] varies by ±50% (Fig. 2d). Although the uniformity assumption no longer holds at 1.8K to legitimize both the vdP-derived resistivity and constraint (2), our procedure to extract the Hall coefficient $R_H$ through the average slope of two Hall resistivity channels is protected by the fundamental principle of voltage-current reciprocity.

With the understanding of how to extract $R_H$ through a $\phi$-dependence study at three fixed temperatures, we now explore the temperature evolution of $R_H$ by taking the average Hall resistivity slope of two reciprocal channels at a fixed $\phi$ (Fig. 2e). $R_H(T)$ measured on two different $Cd_2Os_2O_7$ single crystals (Fig. 3) demonstrates a consistent picture that represents the major finding of this work. $R_H(T)$ remains stable above $T_N$=227K, and all bands only start to evolve at $T_N$, maintaining a delicate balance between them until the first sharp change at 220 K. Despite a rise of nearly three orders of magnitude in $\rho(T)$ from $T_N$ down to 10 K, $R_H(T)$ remains finite and oscillates between positive and negative values. $\rho(T)$ reflects a fast drop in total carrier density with decreasing temperature (Fig. 1), and the oscillating $R_H(T)$ reflects the changes in the bands, with an alternating dominance by either electrons or holes as the itinerant carriers. A temperature dependence of the carrier mobility would not account for this metallic behavior. When a band moves out from the Fermi level, the relative contributions of charge carriers (electron or hole) changes, signaled by a turn in direction in of $R_H(T)$. The first surface moves out from the Fermi level at ~220K (Fig. 3, top inset), and with $R_H(T)$ turning more electron-like, the departed carriers are of hole type. As thermal excitation at finite temperature still populates the departed bands close to the Fermi level, $R_H(T)$ remains a smooth function of $T$. Here we count four characteristic temperatures below $T_N$ – at 220K, 165K, 80K, and 10K – that reflect four sets of carriers leaving the Fermi level (arrows and $T_1$-$T_4$ in Fig. 3 insets). Below $T\sim10$K, a sharply divergent $R_H$ indicates that a full gap emerges in the bulk, and all itinerant carriers are due to thermal excitation, predominately of electron type from the last departed band.

Our results can be compared with existing band structure calculations [23] that suggest the Fermi surface is made up of three sets of carriers in the paramagnetic phase of $Cd_2Os_2O_7$: hole surfaces around the W point, an electron shell at the $\Gamma$ point, and a family of electron ellipsoids along the $\Gamma$-X line, solely of Os $5d$ $t_{2g}$ origin. Our observation reveals a fourth set of hole ellipsoids, potentially located along the $\Gamma$-X line [23], but which have an energy difference too small to be resolved by the band structure calculation given that they drop below the Fermi level at 220K, only 7K below $T_N$. Infrared reflectivity also illuminates the metal-insulator transition in $Cd_2Os_2O_7$ [24-25]. While the direct optical gap has been consistently verified, the interpretation varies from either a spin-density-wave gap opening at $T_N$ [24] or an indirect gap opening at ~210K because of a Liftshitz type of mechanism [25]. The latter is partially consistent with our observed first set of carriers departing the Fermi level at 220K.

Our results demonstrate that the transition at $T_N$=227K in $Cd_2Os_2O_7$ should be regarded as only the magnetic-ordering transition, and there is no concurrent metal-insulator transition despite the fact that the resistivity $\rho(T)$ changes its temperature dependence around $T_N$ (Fig. 1 inset). Given the well-defined changes observed for $R_H(T)$ at $T_N$ (Fig. 3), the driving mechanism behind the electronic evolution should be attributed to Os $5d$ $t_{2g}$ band renormalization by the AIAO order [7], with the effect growing with the increasing strength of the magnetic order parameter, the staggered moment <$m$>, with decreasing $T$. From a direct x-ray magnetic diffraction study [12], <$m$> continues to grow to the zero-temperature limit without saturation. This spin-dependent shift of the quasiparticles' self-energy happens within the antiferromagnetic phase and the electronic gap is opened by *developed* antiferromagnetic order. It is analogous to a Slater mechanism without Brillouin zone-folding [7].

The entire density of states within ±1.5eV of the Fermi level is made of the Os 5$d$ $t_{2g}$ manifold of twelve bands total (two $Cd_2Os_2O_7$ units in the primitive unit cell, with three $t_{2g}$ bands from individual Os ions) [23]. Due to the $Os^{5+}(5d^3)$ valence, these bands are filled between the sixth and seventh bands, with dispersion across the Fermi surface to create the electron-type and hole-type carriers. Spin correlation renormalizes the $t_{2g}$ manifold to create a true gap at the half level between the sixth and seventh bands.

The Mott-Hubbard picture of the metal-insulator transition splits a single band to open a gap by strong charge correlation energy ($U$~2-6eV). In $Cd_2Os_2O_7$, an AIAO antiferromagnet with no structural instability, the transition relies on a small spin-correlation energy ($T_N$~20meV) to essentially create a band insulator. The spin and charge gaps are widely separated in temperature. However, they remain experimentally obscure until the confounding effects of metallic domain wall conduction can be separated from the intrinsic bulk behavior. The methods introduced here clarify the contributions from the bulk and the role of the spin and charge degrees of freedom. They also promise a means to quantify the conductive properties of coercive ferromagnetic domain walls winding their way through an insulating antiferromagnet.

**Methods:**

**Galvanomagnetic measurements:** High quality $Cd_2Os_2O_7$ single crystals were grown by vapor transport techniques [9] and polished from individual tetrahedra to plates of 17-20 μm thickness with a (1, 1, 0) surface normal [14]. The transport samples were selected with a typical mosaic of 0.01°-0.02° FWHM characterized during a previous synchrotron x-ray diffraction study [14]. As the current distribution could potentially change under temperature between domain walls and bulk, a strict Hall bar shaped sample geometry is not useful. Instead, we employ a van der Pauw (vdP) sample geometry [22] to allow the measurement of both Hall and magnetoresistance with their reciprocal configurations (Figs. 1& 2a insets) [21]. Four 25-μm diameter gold wires were attached to the transport samples of typical 200 μm lateral sizes by conductive silver epoxy.

The transport samples were mounted on the standard sample holder of a horizontal rotator probe in a 14-Tesla Physical Property Measurement System (PPMS, Quantum Design, Inc.). Using the horizontal rotation, we first cool our samples through $T_N$ with a 2-Tesla magnetic field applied along an in-plane direction $\phi$ (Fig. 2a insets), then the sample is rotated below $T_N$ to allow the measurement magnetic field of both the Hall effect and the MR to be applied perpendicular to the sample surface. An additional home-built indexing-stage on the sample holder provides degree of freedoms to set field-cooling along 24 discrete angular positions of $\phi$ within the sample surface plane [21]; the origin of the $\phi$-angle has no specific relationship to either the wiring positions of the electrical leads or the crystalline structure. Galvanomagnetic responses of both reciprocal Hall channels and vdP channels were measured at selected temperatures of 195K, 30K, and 1.8K for the full $\phi$-dependence (Fig. 2b-2d). The data set at 195K was measured independently from those at the other two temperatures. At 30K and 1.8K, measurements were repeated at three and eleven $\phi$ positions respectively for a check of reproducibility through an additional field-cooling process. For measurements at 30K, all data except two (one each at $\phi=30°$ and 195°) have a corresponding measurement at 1.8 K during the same field-cooling process in order to compare the $\phi$-dependence at both temperatures. For the full temperature evolution in Figs. 1, 2e, and 3, both samples were field-cooled along one in-plane $\phi$ position to base temperature, then the galvanomagnetic measurements were performed at each stabilized temperature along the warming trajectory. We observe no degrading or change in our samples after many thermal cycles (> 60 in COO-2).

The resistivity was measured using a Lakeshore LS372 AC resistance bridge, working at 9.8 Hz, together with a low-noise 3708 preamp and a home-built vdP switching box based on low-resistance CMOS relay switches. At zero field, the vdP relationship is satisfied to $\Delta R/R_{max}= (R_{Hall1}+R_{vdP1}-R_{vdP2})/Max(R_{vdP1}:R_{vdP2})< \pm 0.05\%$ at 30K and $<\pm 0.25\%$ at 1.8K. $R(H)$ curves were measured over a magnetic field loops of ±4 Tesla. Since no field hysteresis was observed, $R(H)$ values are averaged at each field, exemplified by Fig. 2a, and fit to a polynomial form to the second order. The linear slope is taken for plots in Figs. 2 and 3. Our samples demonstrate a very low level of positive parabolic MR ~0.3% over ±14 Tesla, consistent from 1.65K to 80K, and is thereby unappreciable at low fields.

**Acknowledgments**

Y.F. acknowledges support from the Okinawa Institute of Science and Technology Graduate University, with subsidy funding from the Cabinet Office, Government of Japan. The work at Caltech was supported by National Science Foundation Grant No. DMR-1606858. P.A.L. acknowledges support from the US Department of Energy, Basic Energy Sciences, Grant No. DE-FG02-03ER46076. D.M. acknowledge support from the US Department of Energy, Office of Science, Basic Energy Sciences, Division of Materials Sciences and Engineering.

**Author contributions**

Y.F., Y.W., P.A.L., and T.F.R. conceived of the research; D.M. provided samples; Y.F., Y.W., D.M.S., and S.E.C. performed experiments; Y.F., Y.W., P.A.L., and T.F.R. analyzed data and prepared the manuscript.

**Competing interests**

The authors declare no competing interests.

**References:**

[1] Ma, Y. *et al*. Transparent dense sodium. *Nature* **458**, 182-185 (2009).

[2] Matsuoka, T & Shimizu, K. Direct observation of a pressure-induced metal-to-semiconductor transition in lithium. *Nature* **458**, 186-189 (2009).

[3] Lee, P. A. & Ramakrishnan, T. V. Disordered electronic systems. *Rev. Mod. Phys*. **57**, 287-337 (1985).

[4] Belitz, D. & Kirkpatrick, T.R. The Anderson-Mott transition. *Rev. Mod. Phys*. **66**, 261-380 (1994).

[5] Lee, P. A., Nagaosa, N. & Wen, X. G. Doping a Mott insulator: Physics of high-temperature superconductivity. *Rev. Mod. Phys*. **78**, 17-85 (2006).

[6] Imada, M., Fujimori, A. & Tokura, Y. Metal-insulator transitions. *Rev. Mod. Phys.* **70**, 1039-1263 (1998).

[7] Arita, R., Kunes, J., Kozhevnikov, A. V., Eguiluz, A. G. & Imada, M. *Ab initio* studies on the interplay between spin-orbit interaction and Coulomb correlation in Sr2IrO4 and $Ba_2IrO_4$. *Phys. Rev. Lett.* **108**, 086403 (2012).

[8] Adler, D. Mechanisms for metal-nonmetal transitions in transition-metal oxides and sulfides. *Rev. Mod. Phys*. **40**, 714-736 (1968).

[9] Mandrus, D. *et al*., Continuous metal-insulator transition in the pyrochlore $Cd_2Os_2O_7$, *Phys. Rev. B* **63**, 195104 (2001).

[10] Takatsu, H., Watanabe, K., Goto, K. & Kadowaki, H. Comparative study of low-temperature x-ray diffraction experiments on $R_2Ir_2O_7$ (R = Nd, Eu, and Pr). *Phys. Rev. B* **90**, 235110 (2014).

[11] Wang, Y., Rosenbaum, T. F., Prabhakaran, D., Boothroyd, A.T. & Feng, Y. Approaching the quantum critical point in a highly correlated all-in–all-out antiferromagnet. *Phys. Rev. B* **101**, 220404(R) (2020).

[12] Yamaura, J. *et al*., Tetrahedral magnetic order and the metal-insulator transition in the pyrochlore lattice of $Cd_2Os_2O_7$, *Phys. Rev. Lett*. **108**, 247205 (2012).

[13] Witczak-Krempa, W., Chen, G., Kim, Y. B. & Balents, L. Correlated quantum phenomena in the strong spin-orbit regime. *Annu. Rev. Condens. Matter Phys*. **5**, 57-82 (2014).

[14] Wang, Y. *et al*. Strongly-coupled quantum critical point in an all-in-all-out antiferromagnet. *Nat. Commun*. **9**, 2953 (2018).

[15] Sleight, A. W., Gilison, J. L., Weiher J. F. & Bindloss, W. Semiconductor-metal transition in novel $Cd_2Os_2O_7$. *Solid State Commun*. **14**, 357-359 (1974).

[16] Hiroi, Z., Yamaura, J., Hirose, T., Nagashima, I. & Okamoto, Y. Lifshitz metal–insulator transition induced by the all-in/all-out magnetic order in the pyrochlore oxide $Cd_2Os_2O_7$, *Appl. Phys. Lett. Materials* **3**, 041501 (2015).

[17] Ishikawa, J. J., O'Farrell, E.C.T. & Nakatsuji, S. Continuous transition between antiferromagnetic insulator and paramagnetic metal in the pyrochlore iridate $Eu_2Ir_2O_7$. *Phys. Rev. B* **85**, 245109 (2012).

[18] Sleight, A. W. & Ramirez, A. P. Disappearance of the metal-insulator transition in iridate pyrochlores on approaching the ideal $R_2Ir_2O_7$ stoichiometry. *Solid State. Commun.* **275**, 12-15 (2018).

[19] Néel, L. Some new results on antiferromagnetism and ferromagnetism. *Rev. Mod. Phys.* **25**, 58-63 (1953).

[20] Hirose, H. T., Yamaura, J.-I. & Hiroi, Z. Robust ferromagnetism carried by antiferromagnetic domain walls. *Sci. Rep.* **7**, 42440 (2017).

[21] Wang, Y. *et al*., Antisymmetric linear magnetoresistance and the planar Hall effect. *Nat. Commun*. **11**, 216 (2020).

[22] van der Pauw, L. J. A method of measuring the resistivity and Hall coefficient on lamellae or arbitrary shape. *Philips Tech. Rev*. **20**, 220-224 (1958).

[23] Singh, D. J., Blaha, P., Schwarz, K. & Sofo, J. O. Electronic structure of the pyrochlore metals $Cd_2Os_2O_7$ and $Cd_2Re_2O_7$. *Phys. Rev. B* **65**, 155109 (2002).

[24] Padilla, W. J., Mandrus, D. & Basov, D. N. Searching for the Slater transition in the pyrochlore $Cd_2Os_2O_7$ with infrared spectroscopy. *Phys. Rev. B* **66**, 035120 (2002).

[25] Sohn, C. H. *et al*. Optical spectroscopic studies of the metal-insulator transition driven by all-in–all-out magnetic ordering in 5*d* pyrochlore $Cd_2Os_2O_7$. *Phys. Rev. Lett.* **115**, 266402 (2015).

**Figure captions:**

**Fig. 1. Transport signatures of $Cd_2Os_2O_7$ at zero field.** Resistivity $\rho(T)$ of $Cd_2Os_2O_7$ measured over two single crystals (COO-1 and COO-2) using the vdP configuration (Schematics). At each temperature, both vdP channels were measured in order to account for the changing current path between the insulating bulk and conductive domain walls. (Top inset) The resistance ratio $R_{vdP2}/R_{vdP1}$ between two vdP channels manifests the changing current path below $T_N$=227K. (bottom inset) Details of $\rho(T)$ near $T_N$, showing both metallic behavior above $T_N$, and also an upturn of $\rho$ near but still above $T_N$, which is often attributed to dynamical spin fluctuation effects without long-range order. $T_N$ is determined precisely from magnetic susceptibility $\chi(T)$, individual $R_{vdP2}(T)$ and $R_{vdP1}(T)$ (not shown), and $R_H(T)$ in Fig. 3.

**Fig. 2. Separating the influence of metallic ferromagnetic domain walls.** (**a**) Raw data of Hall resistances $R(H)$ between two reciprocal channels. The difference in slopes indicates the influence of asymmetric linear magnetoresistance from ferromagnetic domain walls [21]. (Top schematics) The field-cooling process in two-stages: (i) field aligned parallel to the sample surface during cooldown; and (ii) sample rotated to have field perpendicular to its surface for galvanomagnetic measurements. The in-plane magnetizing direction is defined by angle $\phi$ at 24 discrete positions. (Bottom schematics) vdP configuration for the two reciprocal Hall channels. (**b-d**) $\phi$-dependence of Hall resistivity slopes in two reciprocal channels (red and blue) and their average (fresh green), in addition to resistivity (green) and vdP ratio (orange), measured in sample COO-2 at three temperatures 195K, 30K, and 1.8K. (Methods) (**e**) Hall resistivity slopes from two reciprocal channels (red and blue) are plotted alongside their average (fresh green) for two

different single crystal $Cd_2Os_2O_7$ samples. Although the resistance slopes of an individual Hall channel are very different for each sample, with occasional crossings at various temperatures, the averages are similar in shape between the two samples, and determine the bulk Hall coefficient $R_H(T)$. This irregular behavior of resistance slopes of individual Hall channel highlights potential experimental deficiencies in previous Hall measurements.

**Fig. 3. Metal-insulator transition in $Cd_2Os_2O_7$ revealed through bulk Hall coefficient.** The bulk Hall coefficient $R_H(T)$ from two $Cd_2Os_2O_7$ samples in Fig. 2e are compared in detail. While $R_H(T)$ evolves slowly above $T_N$, it starts to deviate from the high temperature behavior right at $T_N$, indicating that the onset of AIAO antiferromagnetic order influences the band structure around the Fermi level. Nevertheless, $R_H(T)$ does not diverge until it reaches a temperature below 10K ~ 0.044 $T_N$, when the true charge gap opens. In between, each sharp change in in $R_H(T)$ indicates a band leaving the Fermi level, marked by arrows at four different temperatures, $T_1$=220K, $T_2$=165K, $T_3$=80K, and $T_4$=10K. Two bands are of hole type and the other two of electron type.

Fig. 1.

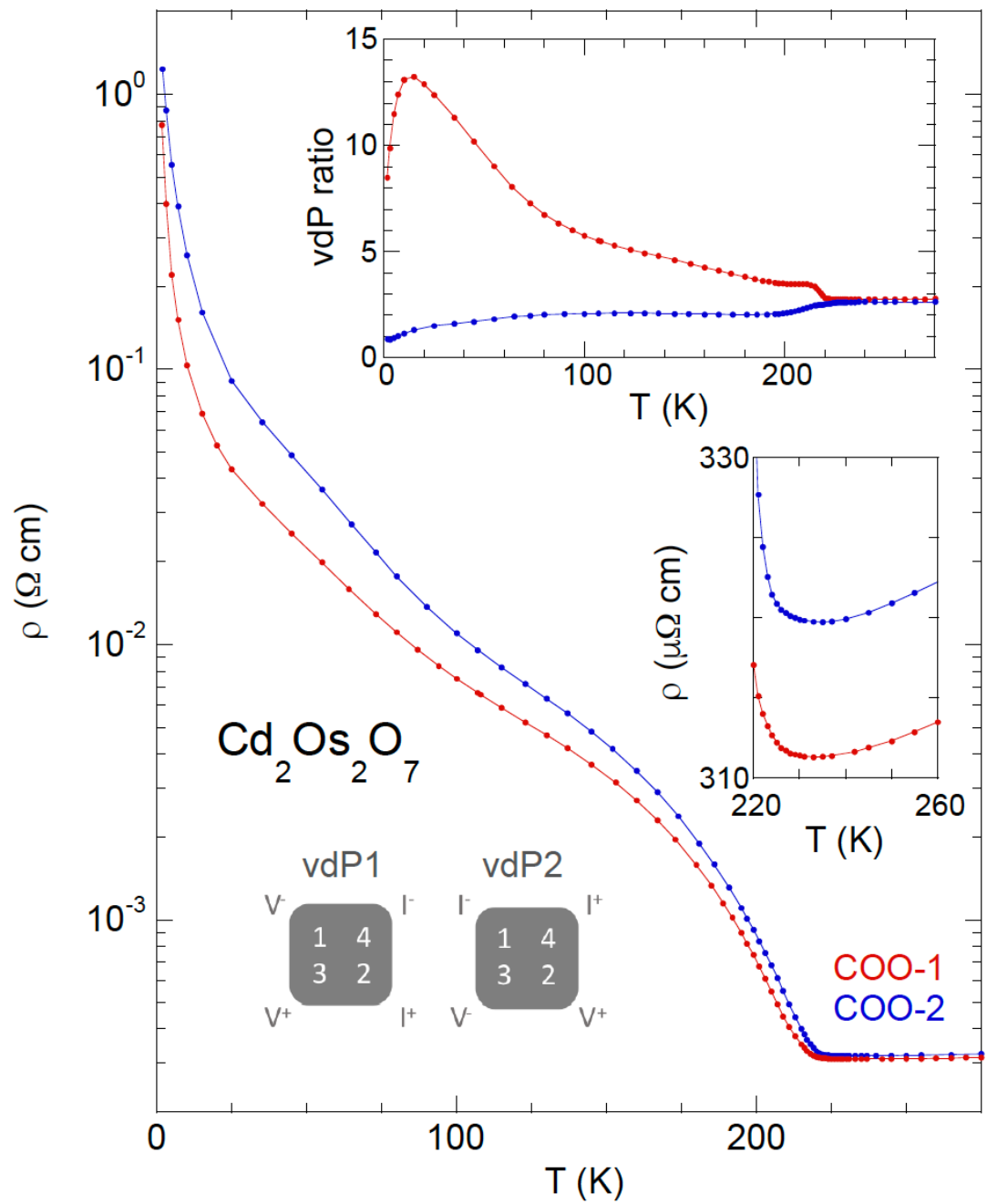

Fig. 2.

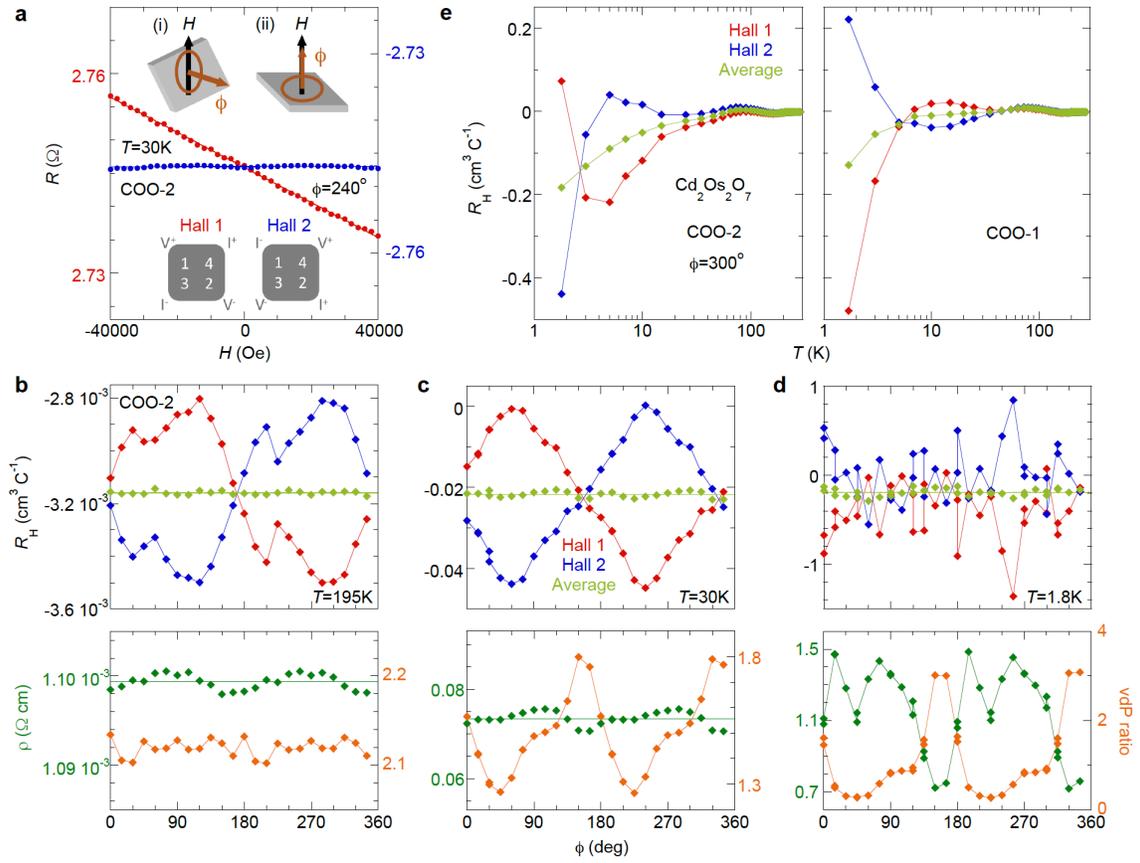

Fig. 3.

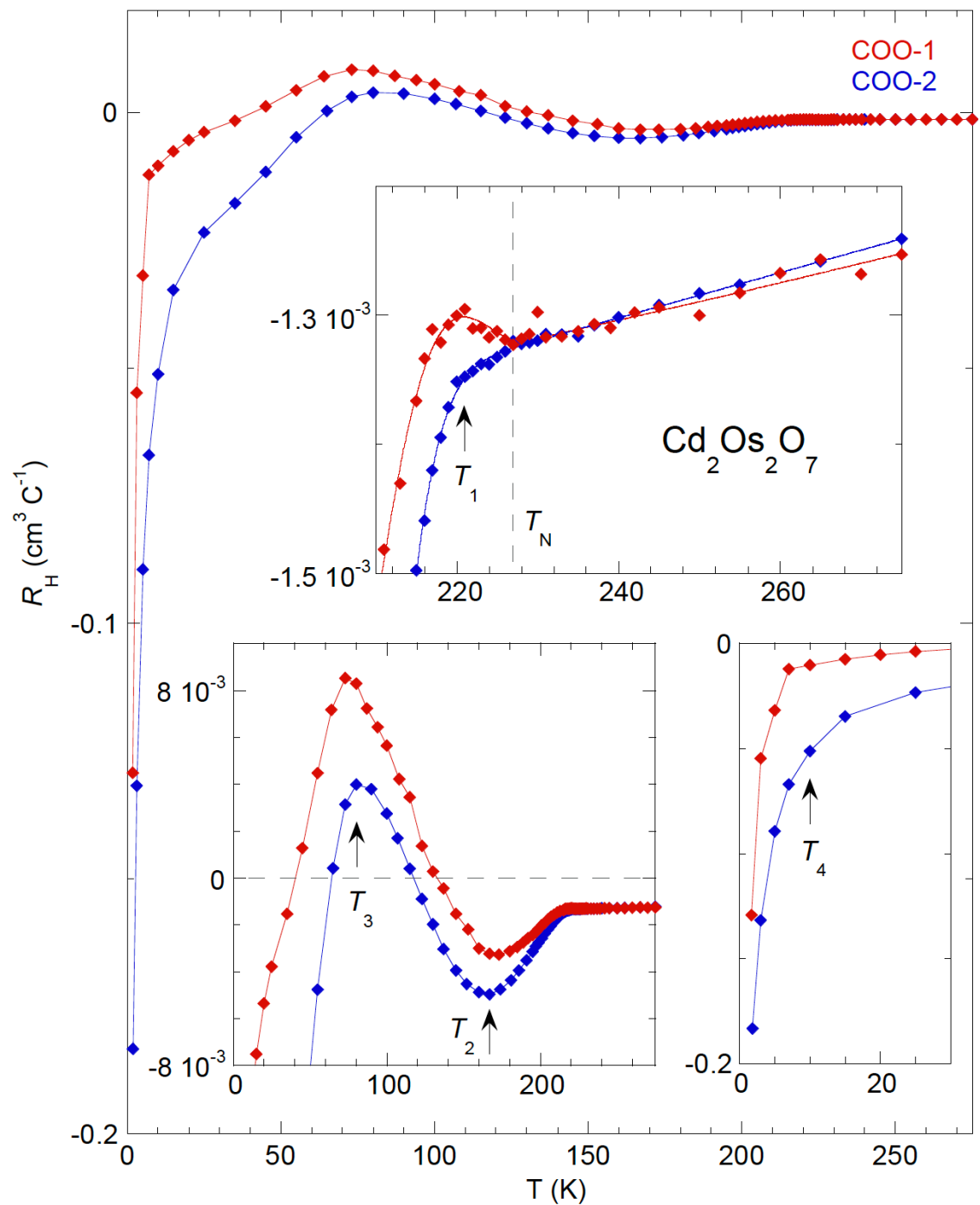